\documentclass[aps,prb,twocolumn,bibnotes,longbibliography,superscriptaddress]{revtex4-1}
\usepackage{t1enc}
\usepackage[utf8]{inputenc}
\usepackage{amsmath}
\usepackage{graphicx}

\renewcommand{\Re}{\textrm{Re}}

\begin{document}

\title{Broadband microwave spectroscopy of semiconductor nanowire-based
Cooper-pair transistors}

\author{Alex Proutski}
\thanks{These authors contributed equally to this work.}

\author{Dominique Laroche}
\thanks{These authors contributed equally to this work.}

\author{Bas van 't Hooft}

\affiliation{QuTech and Kavli Institute of Nanoscience, Delft University of
Technology, 2600 GA Delft, The Netherlands}

\author{Peter Krogstrup}

\affiliation{Microsoft Quantum Materials Lab Copenhagen, Niels Bohr Institute,
University of Copenhagen, 2100 Copenhagen, Denmark}

\affiliation{Center for Quantum Devices, Niels Bohr Institute,
University of Copenhagen, 2100 Copenhagen, Denmark}

\author{Jesper Nyg{\aa}rd}

\affiliation{Center for Quantum Devices, Niels Bohr Institute,
University of Copenhagen, 2100 Copenhagen, Denmark}

\author{Leo P. Kouwenhoven}

\affiliation{Microsoft Quantum Lab Delft, 2600 GA Delft, The Netherlands}

\affiliation{QuTech and Kavli Institute of Nanoscience, Delft University of
Technology, 2600 GA Delft, The Netherlands}

\author{Attila Geresdi}
\email[Corresponding author; e-mail address: ]{a.geresdi@tudelft.nl}

\affiliation{QuTech and Kavli Institute of Nanoscience, Delft University of
Technology, 2600 GA Delft, The Netherlands}

\date{\today}

\begin{abstract}
The Cooper-pair transistor (CPT), a small superconducting island enclosed
between two Josephson weak links, is the atomic building block of various
superconducting quantum circuits. Utilizing gate-tunable semiconductor channels
as weak links, the energy scale associated with the Josephson tunneling can be
changed with respect to the charging energy of the island, tuning the extent of
its charge fluctuations. Here, we directly demonstrate this control by mapping
the energy level structure of a CPT made of an indium arsenide nanowire
(NW) with a superconducting aluminum shell. We extract the device
parameters based on the exhaustive modeling of the quantum dynamics of the
phase-biased nanowire CPT and directly measure the even-odd parity occupation
ratio as a function of the device temperature, relevant for superconducting and
prospective topological qubits.
\end{abstract}

\maketitle

The energy landscape of a Cooper-pair transistor (CPT), a mesoscopic
superconducting island coupled to superconducting leads via two Josephson
junctions, is determined by the interplay of the electrostatic addition energy
of a single Cooper pair, $E_\textrm{C}=(2e)^2/2C$ \cite{Averin_1991, Grabert_1991}, and
the coherent tunneling of Cooper pairs, characterized by the Josephson energy $E_\textrm{J}$
\cite{Josephon_1962, Ambegaokar1963}.

The electronic transport through CPTs has mostly been studied for metallic
superconducting islands enclosed between tunnel junctions by voltage bias
spectroscopy \cite{PhysRevLett.59.109, PhysRevLett.65.377, Tuominen_1992},
switching current measurements \cite{Joyez1994, PhysRevB.50.627, Aumentado_2004,
Woerkom2015}, microwave reflectometry \cite{Ferguson_2006,PhysRevB.78.024503},
 and broadband microwave spectroscopy \cite{Billangeon_2007,
 PhysRevLett.98.216802}.
Recent material developments \cite{Krogstrup2015, Gazibegovic2017} made it
possible to investigate superconducting transport in semiconductor nanowire (NW) weak
links, which lead to Andreev level quantum circuits \cite{Woerkom_2017_ABS,
PhysRevLett.121.047001, PhysRevX.9.011010} and gate-tunable superconducting
quantum devices
\cite{PhysRevLett.115.127001,PhysRevLett.115.127002,PhysRevLett.120.100502,
PhysRevB.99.085434}. In addition, hybrid superconductor-semiconductor island
devices, which are the atomic building blocks of proposed topological quantum
bits based on Majorana zero-energy modes
\cite{PhysRevB.88.035121,Aasen_2016,PhysRevB.95.235305,Plugge2017}, have been
fabricated and measured using normal metallic leads \cite{Albrecht_2016,
Shen2018}, but thus far there is very limited experimental work on hybrid CPTs
with superconducting leads \cite{PhysRevB.98.174502}.

Such applications require the control of the Josephson coupling via the
semiconductor weak link \cite{PhysRevLett.119.187704}. In addition, the charging
energy of a NW CPT can deviate from the predictions of the orthodox theory
\cite{Averin_1991, Grabert_1991} due to renormalization effects arising because
of finite channel transmissions \cite{PhysRevLett.82.3685}. Therefore,
understanding the quantum dynamics of CPTs with semiconductor weak links is
crucial for these hybrid device architectures.

Here we directly measure the transitions between the energy levels of a NW CPT.
The CPT is embedded in the circuit shown in Fig.~1(a). The superconducting island is created
from an indium arsenide (InAs) nanowire with an epitaxial layer of aluminium
(Al) \cite{Krogstrup2015} between two Josephson junctions, formed
by removing two sections of the Al shell with a wet chemical etch. We investigated two devices,
both with $100\,$nm long junctions and island lengths of $800\,$nm and
$1.75\,\mu$m for device 1 and device 2 [enclosed in the red box in Fig.~1(a)],
respectively. The junctions are tuned via their respective local electrostatic
gates, $V_\textrm{tg1}$ and $V_\textrm{tg2}$. The gate charge,
$n_\textrm{g}=V_\textrm{g} C_\textrm{g}/2e$, is set by the gate voltage
$V_\textrm{g}$ and the effective gate capacitance, $C_\textrm{g}$ (see the right panel in Fig.~1(a) and
the supplementary information \cite{supplement}).
The nanowire CPT is embedded in a superconducting quantum interference device (SQUID) with an
Al/AlO$_x$/Al tunnel junction [in the yellow box in Fig.~1(a)] which exhibits a
much higher Josephson energy than the CPT. This asymmetry ensures that the applied
phase $\varphi=2\pi\Phi/\Phi_0$ drops mostly over the CPT. Here,
$\Phi$ is the applied flux and $\Phi_0=h/2e$ is the superconducting flux
quantum.

\begin{figure}[!ht]
\includegraphics[width=0.5\textwidth]{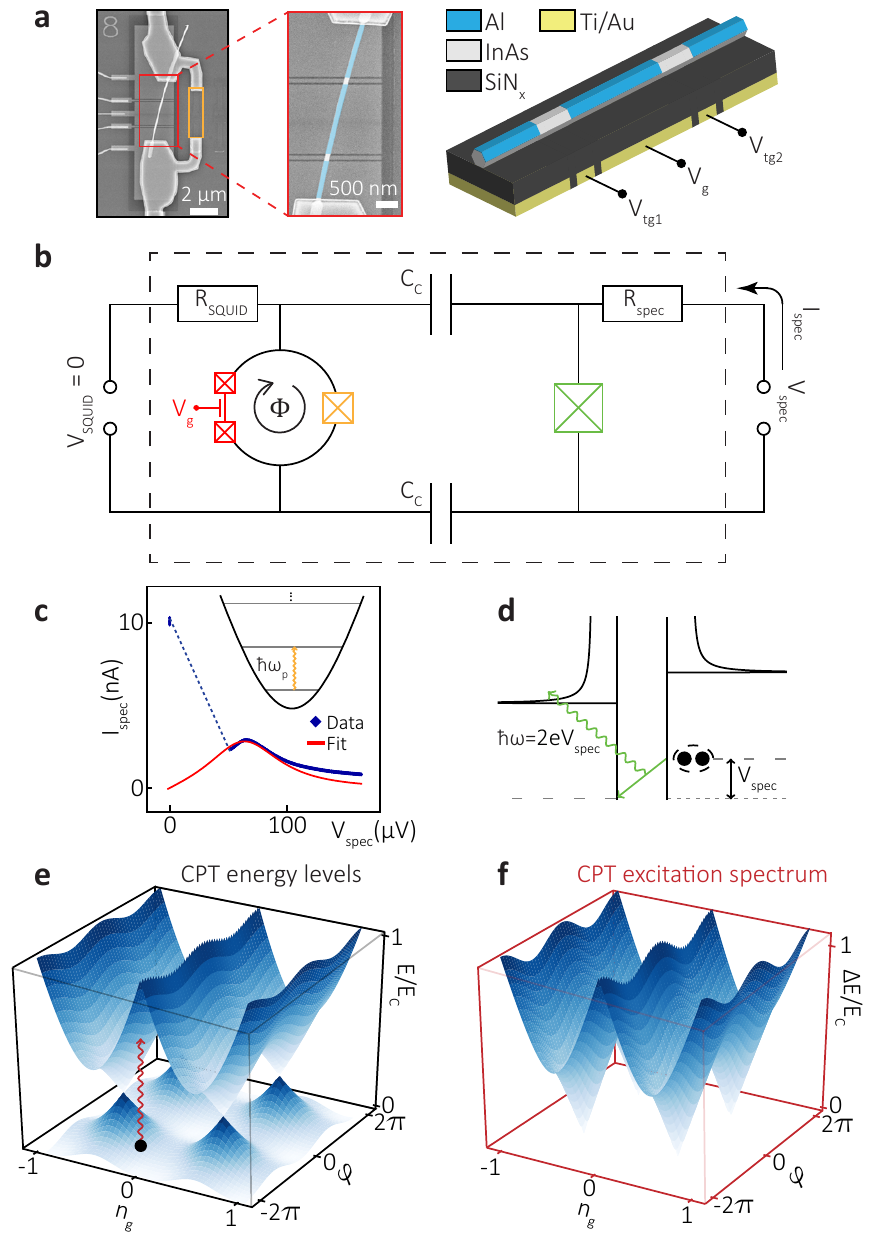}
\caption{(a) Left: Scanning electron micrograph of the nanowire CPT (in red
box) and an Al/AlO$_x$/Al tunnel junction (yellow box) forming the hybrid SQUID
loop. Middle: False colored micrograph of the nanowire CPT (device 2). Right:
Three dimensional sketch of the CPT on the three electrostatic gates. (b) Equivalent
circuit schematics with the hybrid SQUID on the left and a single
Al/AlO$_x$/Al tunnel junction used as a spectrometer (green box) on the
right side. The circuit elements within the black dashed box are on-chip and
cooled to $T \sim 18\,$mK. (c) $I(V)$ trace of the spectrometer with the CPT arm
in full depletion (device 1). The red solid line shows the fit to the circuit
model of a single resonance centered at $\hbar \omega_\textrm{p}=148\,\mu$eV driven by the photons
with an energy of $\hbar \omega = 2eV_\textrm{spec}$ emitted by the
spectrometer junction (d). The calculated energy bands (e) and
transition energies (f) of a CPT with $E_\textrm{J1}=E_\textrm{J2}=E_\textrm{C}/4$ as a function of
gate charge, $n_\textrm{g}$ and total phase bias, $\varphi$.}
\end{figure}

We utilized a capacitively coupled  Al/AlO$_x$/Al superconducting tunnel
junction as a broadband on-chip microwave spectrometer [green box in Fig.~1(b)]
\cite{Billangeon_2007, Bretheau_2013_Andreev, Woerkom_2017_ABS}, where inelastic
Cooper-pair tunneling gives rise to a dc current contribution in a dissipative
environment \cite{Holst_1994}:

\begin{equation}\label{eq.1}
I_\textrm{spec} =
\frac{I_\textrm{c,spec}^{2}\Re[Z\left(\omega\right)]}{2V_\textrm{spec}}.
\end{equation}      

Here, $I_\textrm{c,spec}$ is the critical current of the spectrometer tunnel
junction and $Z\left(\omega\right)$ is the impedance of the environment at the
frequency $\omega = 2eV_\textrm{spec}/\hbar$, determined by the spectrometer dc
voltage bias, $V_\textrm{spec}$ [Fig.~1(d)]. This dc to microwave conversion
allowed us to directly measure the excitation energies of the hybrid SQUID,
where $\textrm{Re}[Z\left(\omega\right)]$ exhibits a local maximum
\cite{Kos_2013}. To reduce microwave leakage, we applied the bias voltages to
the hybrid SQUID and to the spectrometer junctions via on-chip resistors,
yielding $R_\textrm{SQUID}=12\,\textrm{k}\Omega$ and
$R_\textrm{spec}=2.8\,\textrm{k}\Omega$. The chip [in black dashed box in
Fig.~1(b)] was thermally anchored to the mixing chamber of the dilution
refrigerator with a base temperature of $\approx$ $18\,$mK. Full details of the
fabrication process and device geometry are given in the supplementary
information \cite{supplement}.

\begin{figure}
\centering
\includegraphics[width=0.5\textwidth]{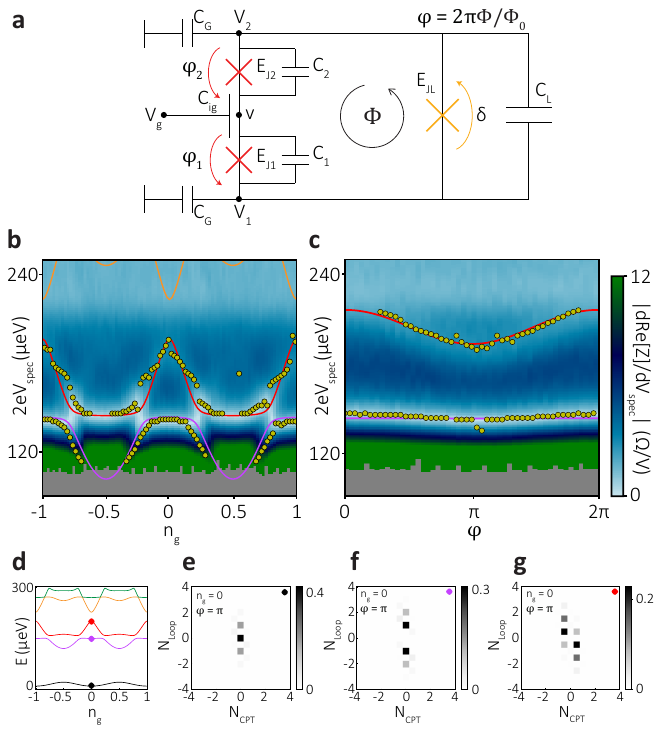}
\caption{(a) Equivalent schematics of the hybrid SQUID used to build the circuit
Hamiltonian. Observed transitions for device 1 as a function of the gate charge
$n_\textrm{g}$ at a fixed $\varphi=\pi$ (b) and applied phase bias $\varphi=2\pi
\Phi/\Phi_0$ at a fixed $n_\textrm{g}=0$ (c).
The transitions are identified at the local minima of $|d\Re(Z)/dV_\textrm{spec}|$ (yellow dots).
The best fit is shown as solid lines, yielding $E_\textrm{C1} = 168\,\mu$eV, $E_\textrm{C2} =
260\,\mu$eV, $E_\textrm{Cc} = 188\,\mu$eV, $E_\textrm{J1} = 132\,\mu$eV and  $E_\textrm{J2}
= 16\,\mu$eV, see text. (d) The corresponding energy bands of the device as a
function of $n_\textrm{g}$ at $\varphi=\pi$. The two-component probability distributions
of the ground state (e), first excited state (f) and second excited state (g) at
$n_\textrm{g}=0$ and $\varphi=\pi$, denoted by circles of the corresponding color in panel
(d) (see text). See Fig.~3 upper row for gate voltage values.}
\end{figure}

We begin by analyzing the circuit while keeping both nanowire junctions in full
depletion by applying large negative gate voltages $V_\textrm{tg1}$ and
$V_\textrm{tg2}$.
The $I(V)$ curve of the spectrometer of device 1 is shown in Fig.~1(c). A clear
peak is observed with an amplitude of $3\,$nA centered at $\approx 75\,\mu$eV.
We attribute this peak to the plasma resonance of the tunnel junction in the
SQUID at $\hbar\omega_\textrm{p}=\sqrt{2E_\textrm{JL}E_\textrm{CL}}$.
Here $E_\textrm{JL}=\Delta_\textrm{J}h/(8e^2R_\textrm{J})=249\,\mu$eV is the Josephson energy
\cite{Ambegaokar1963}, with $\Delta_\textrm{J}=245\,\mu$eV being the 
measured superconducting gap and $R_\textrm{J}=3.17\,\textrm{k}\Omega$ the normal state
resistance of the junction, acquired at a voltage bias much higher than
$2\Delta_\textrm{J}$.
This value yields $E_\textrm{CL}=2e^2/C_\textrm{L}=44\,\mu$eV and a shunt capacitance
$C_\textrm{L}=7.28\,$fF. Fitting the resonant peak using Eq.~(1), we find a quality
factor $Q\approx1$ and a characteristic impedance $Z_0=610\,\Omega\ll
R_\textrm{q}=h/4e^2$, which together ensure the validity of Eq.~(1) describing a direct
correspondence between the measured $I_\textrm{spec}$ and $\Re[Z(\omega)]$. We
note that we found very similar values for device 2 as well (see supplementary
information for a detailed analysis and a list of parameters \cite{supplement}).

\begin{figure*}[ht!]
	\centering
	\includegraphics[width=0.9\textwidth]{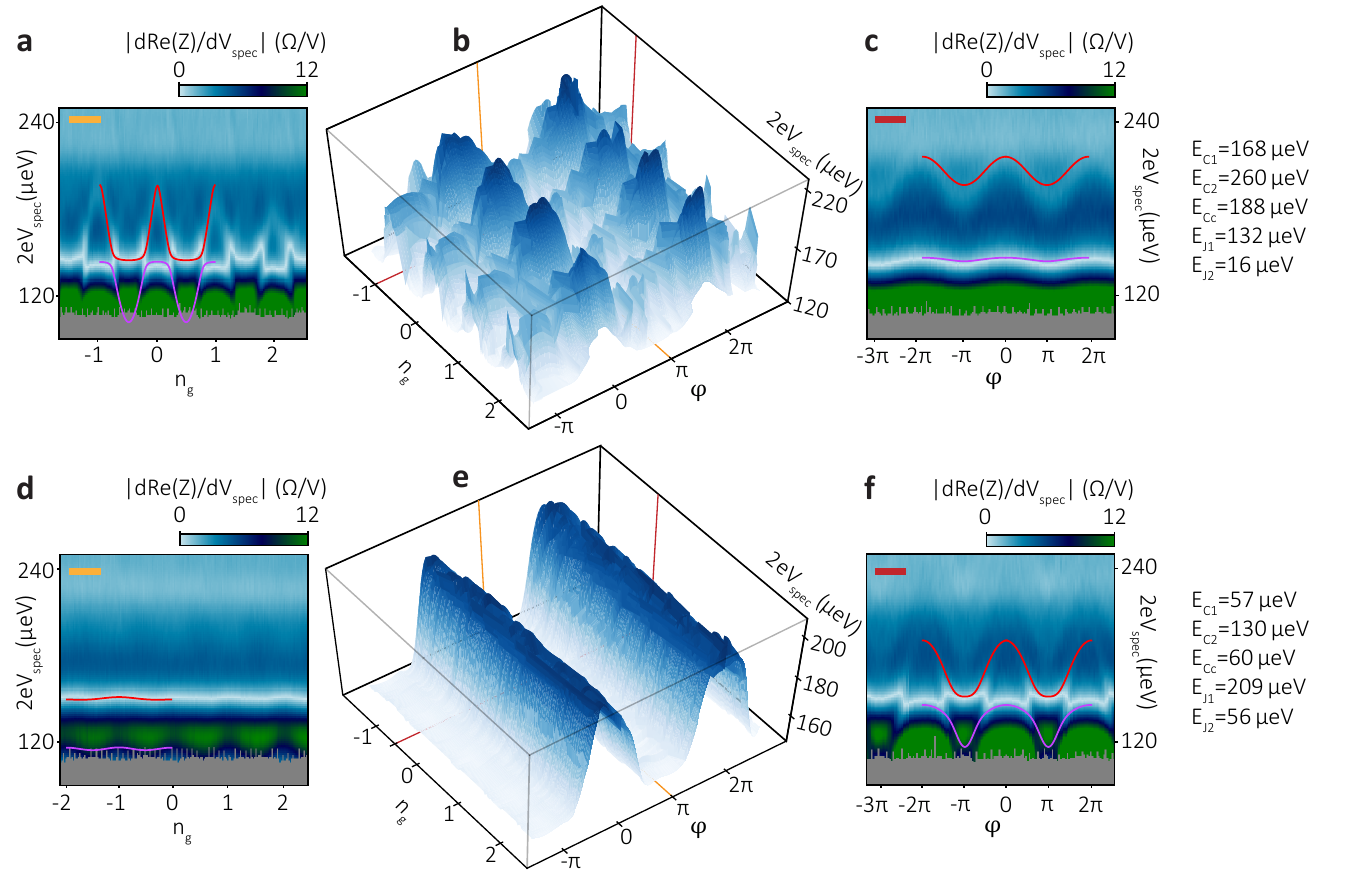}
	\caption{Upper row: The measured excitations spectrum of device 1 as a function
	of $n_\textrm{g}$ and $\varphi$ with $V_\textrm{tg1} = 0.55\,$V, $V_\textrm{tg2} =
	0.85\,$V and $V_\textrm{g}=250\ldots264.5\,$mV, same as in Fig.~2. (b)
	shows the full map of the second excitation whereas the linecut data are shown at the positions denoted
	by the orange and red lines respectively in (a) and (c).
	Bottom row: Measured data on the same device with $V_\textrm{tg1} = 1.5\,$V,
	$V_\textrm{tg2} = 1.545\,$V and $V_\textrm{g}=916.9\ldots931.4\,$mV. Note the
	weak dependence on $n_\textrm{g}$ due to the more open semiconductor channels. The best fits
	of the first two excitation energies are overlain in (a), (c) and (d),
	(e). All data were taken on device 1, the parameters of the best fit are listed
	on the right for each setting.}
\end{figure*}

Next, we investigate the spectrometer response to the applied gate voltage
$V_\textrm{g}$ and phase $\varphi$ [Fig.~2(b) and (c)] when the Josephson junctions
are opened by setting positive gate voltages $V_\textrm{tg1}$ and $V_\textrm{tg2}$.
The excitations of the CPT are superimposed on that of the plasma resonance, so
we display $|d\Re(Z)/dV_\textrm{spec}|$ to reach a better visibility of the
transitions (see the supplementary information for a comparison
\cite{supplement}). Note that we show the excitation energy $\hbar\omega=2eV_\textrm{spec}$ on the vertical axis for all spectra.
This measurement yields clear oscillations as a function of both $n_\textrm{g}$ and
$\varphi$, consistent with the expected periodic behavior of the CPT energy
levels \cite{Joyez1994}. We note that the finite load resistance of the
spectrometer $R_\textrm{spec}$ prevented us from measuring the transitions below
$2eV_\textrm{spec}=103\,\mu$eV.

We model our device with the schematics depicted in
Fig.~2(a) and build the Hamiltonian of the circuit based on conventional
quantization procedures \cite{PhysRevA.69.062320, doi:10.1002/cta.2359}. We
use the conjugate charge and phase operators which pairwise obey
$[\hat{\varphi}_{1,2},\hat{N}_{1,2}]=i$ and note that
$\hat{\delta}=\varphi-\hat{\varphi}_1-\hat{\varphi}_2$:
\begin{multline}
\hat{H} = \frac{1}{2}E_\textrm{C1}(\hat{N}_1-n_\textrm{g})^{2} +
\frac{1}{2}E_\textrm{C2}(\hat{N}_2+n_\textrm{g})^{2}\\
-  \frac{1}{2}E_\textrm{Cc}(\hat{N}_1-n_\textrm{g})(\hat{N}_2+n_\textrm{g}) \\
- E_\textrm{J1} \cos(\hat{\varphi}_1) - E_\textrm{J2} \cos(\hat{\varphi}_2) - E_\textrm{JL}
\cos(\varphi - \hat{\varphi}_1 - \hat{\varphi}_2).
\label{eq:7}
\end{multline}

Here, the charging of the circuit is described by the effective parameters
$E_\textrm{C1}$, $E_\textrm{C2}$ and $E_\textrm{Cc}$ set by the capacitance values $C_1$, $C_2$,
$C_\textrm{L}$, $C_\textrm{ig}$ and $C_\textrm{G}$ with a functional form provided in the supplementary
information \cite{supplement}. The Cooper-pair tunneling is characterized by the
Josephson energies of the three junctions, $E_\textrm{J1}$, $E_\textrm{J2}$ and
$E_\textrm{JL}$, respectively. We note that we set $E_\textrm{JL}=249\,\mu$eV
for the analysis below.

To calculate the excitation spectrum, we solve the eigenvalue problem 
to find $E_i(n_\textrm{g},\varphi)$, where $\hat{H}\Psi_i=E_i\Psi_i$, and compute the
transition energies $\hbar \omega_i=E_i-E_0$, with $E_0$ being the ground state
energy of the system. This model allows us to fit the excitation spectra
simultaneously as a function of $n_\textrm{g}$ and $\varphi$ based on the first
two transitions (red and purple solid lines for $\hbar\omega_1$ and
$\hbar\omega_2$, respectively) against the measured data (yellow circles in
Fig.~2). For illustration, we also display $\hbar\omega_3$ (orange line) in
Fig.~2(b) using the same fit parameters, however, this transition was not observed in the
experiment.

To understand the nature of the excited levels, we calculate the energy bands of
the hybrid SQUID using the fitted parameters [Fig.~2(d)] and evaluate the probability
distribution $p_i(N_1,N_2)=|\Psi_i(N_1,N_2)|^2$, where $N_1$ and $N_2$ form the
charge computational basis. However, it is more instructive to use the charge
numbers $N_\textrm{CPT}=N_1-N_2$ and $N_\textrm{Loop}=N_1+N_2$.
Intuitively, $N_\textrm{CPT}$ and $N_\textrm{Loop}$ represent the excess number
of Cooper pairs on the island and in the loop, respectively. Indeed, the ground
state wavefunction is centered around $N_\textrm{CPT}=N_\textrm{Loop}=0$
[Fig.~2(e)]. Conversely, the probability distribution of the first excited state
[Fig.~2(f)] exhibits a bimodal distribution in $N_\textrm{Loop}$, consistently with the
first plasma mode excitation but no excess charge on the CPT [purple circle in
Fig.~2(d)]. This is in contrast with the wavefunction of the next energy level
[Fig.~2(g) and red circle in Fig.~2(d)], which is centered around
$N_\textrm{CPT}=\pm1$. This analysis demonstrates the coupling between the
plasma and localized charge degrees of freedom \cite{Wallraff_2004}.

Next, we investigate the impact of $V_\textrm{tg1}$ and $V_\textrm{tg2}$
on the CPT spectrum. In Fig.~3, we show the measured spectra for two distinct
gate settings. Remarkably, almost a full suppression of the charge dispersion is
achieved by an $\approx 1\,$V increase in $V_\textrm{tg1}$ and $V_\textrm{tg2}$,
showcasing the feasibility of topological quantum bit designs relying on the modulation of the
charge dispersion in superconductor-semiconductor hybrid devices
\cite{Aasen_2016}. Furthermore, we observe a strong renormalization of the
characteristic charging energies in the open regime \cite{PhysRevLett.82.3685,
PhysRevLett.122.016801}, which does not exist for the case of fully metallic
CPTs with tunnel junctions, where the charging energy is fully determined by the
device geometry. In addition, we find an increase in the Josephson
energies $E_\textrm{J1,2}$, further contributing to the suppression of the
charge dispersion of the CPT in the limit of $E_\textrm{J} \gg E_\textrm{C}$
\cite{PhysRevA.76.042319}.

Thus far, we only considered the even charge occupation of the island, where all
electrons are part of the Cooper-pair condensate, and a single quasiparticle
occupation is exponentially suppressed in $\Delta/k_\textrm{B} T$, where $\Delta$ is the
superconducting gap \cite{Averin1992}. However, a residual odd population is
typically observed in the experiments, attributed to a non-thermal quasiparticle
population in the superconducting circuit. In our experiment, we also find an
additional spectral line, shifted by $\delta n_\textrm{g}=0.5$ [see Figs 2(b) and 3(a)],
substantiating a finite odd number population of the island. We investigate this
effect as a function of the temperature, and find that above a typical
temperature of $T^\star\approx 300\,$mK, the measured signal is fully $1e$
periodic [Fig.~4(b)], in contrast to the $2e$ periodic data taken at $18\,$mK
[Fig.~4(a)].

To quantify the probability of the even and odd occupations, we extract the
gate-charge dependent component of the measured spectra $\delta
I_\textrm{spec}(n_\textrm{g})$ to evaluate $\delta I_\textrm{odd}=\delta
I_\textrm{spec}(n_\textrm{g}=0.5)$ and $\delta I_\textrm{even}=\delta
I_\textrm{spec}(n_\textrm{g}=0)$, see the inset in Fig.~4(c). We now make the assumption
that the microwave photon frequency is much higher than the parity switching
rate of the CPT. We evaluate the current response at
$hf=2eV_\textrm{spec}=180\,\mu$eV [see Figs.~4(a) and (b)] corresponding to
$f=43.5\,$GHz, well exceeding parity switching rates measured earlier on similar
devices \cite{Albrecht_2017, PhysRevB.98.174502}. In this limit, the
time-averaged spectrometer response is the linear combination of the signals
corresponding to the two parity states and $\delta I_\textrm{even,odd}\sim
p_\textrm{even,odd}$, respectively. From this linear proportionality,
$p_\textrm{even}=(1+\delta I_\textrm{odd}/\delta I_\textrm{even})^{-1}$ follows.

We plot the extracted $p_\textrm{even}$ in Fig.~4(c). We find that above a
crossover temperature $T^\star\approx 300\,$mK, $p_\textrm{even}$ approaches
$1/2$, in agreement with the commonly observed breakdown of the parity effect at
$T^\star\ <\Delta$ as a result of the vanishing even-odd free energy difference
\cite{Lafarge_1993,Woerkom2015,Higginbotham2015},
\begin{equation}
\Delta F = -k_\textrm{B}T\ln\tanh \left(N_\textrm{eff}e^{-\Delta/k_\textrm{B} T} \right).
\end{equation}
Here, $N_\textrm{eff} = \rho V \sqrt{2\pi k_\textrm{B} T \Delta}$ at a temperature of
$T$ with the island volume being $V$. We use the density of states at the Fermi
level in the normal state $\rho=1.45\times
10^{47}\,\textrm{J}^{-1}\textrm{m}^{-3}$ for aluminium \cite{Ferguson_2006}.
Then the even charge parity occupation is given by
$p_\textrm{even} = 1-1/\left(1 + e^{\Delta F/k_\textrm{B}T}\right)$.

While this analysis describes the breakdown of the even-odd effect [see
blue dashed line as the best fit in Fig.~4(c)], it fails to account for the
observed saturation $p_\textrm{even}\sim 0.8<1$ in the low temperature limit, at
$T<150\,$mK. This saturation can be be phenomenologically understood based on a
spurious overheating of the island. We assume that the electron temperature $T_e =
\left(T_0^5 + T^{5}\right)^{1/5}$, where the chip (phonon) temperature is $T$,
and the electron saturation temperature is $T_0$ due to overheating and weak
electron-phonon coupling at low temperatures \cite{Giazotto2006}.

The resulting best fit is shown as a solid red line in Fig.~4(c). We find a
metallic volume of $V=4.66\times10^{-23}\,\textrm{m}^3$, consistent with the
micrograph shown in Fig.~1(a). The fit yields a superconducting gap
$\Delta=140\pm3\,\mu$eV, slightly lower than the that of bulk aluminum, which is
expected due to the presence of induced superconductivity in the semiconductor.
The fitted saturation temperature $T_0=244\pm4\,$mK and limiting
$p_\textrm{odd}=1-p_\textrm{even}\approx 0.17$ demonstrates the abundance of
non-equilibrium quasiparticles, in agreement with recent experimental findings
\cite{PhysRevLett.121.157701, mannila2018parity} on metallic devices.
However, the unpaired quasiparticle density \cite{Higginbotham2015}
$n_\textrm{qp}=V^{-1}N_\textrm{eff}^2\exp(-2\Delta/k_\textrm{B} T_0)\approx800\,\mu$m$^{-3}$ is
orders of magnitude higher than typical values for all-metallic devices, falling
in the range $n_\textrm{qp}\approx0.01\ldots1\,\mu$m$^{-3}$
\cite{Ferguson_2006,Woerkom2015,mannila2018parity}. The same analysis was also
performed on device 1 yielding similar results, see the supplementary
information \cite{supplement}. Our results substantiate the importance of
controlling the quasiparticle population for hybrid semiconductor-superconductor CPTs in
prospective topological quantum bits to decrease their rate of decoherence
\cite{PhysRevB.85.174533}.

\begin{figure}
	\centering
	\includegraphics[width=0.5\textwidth]{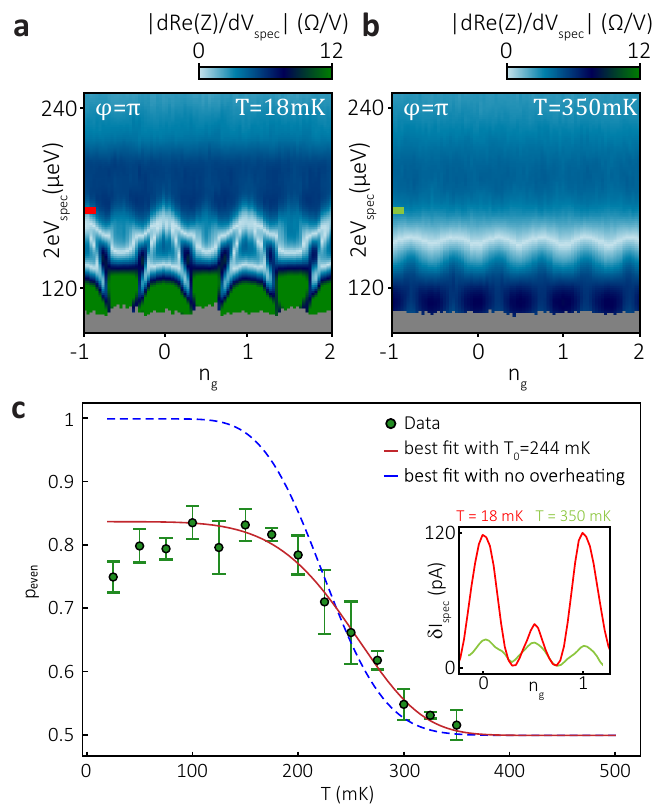}
	\caption{The measured excitation spectra of device 2 at $V_\textrm{tg1} =
	1.4\,$V, $V_\textrm{tg2} = 1.5025\,$V as a function of $n_\textrm{g}$ corresponding to
	$V_\textrm{g}=-533\ldots-527.8\,$mV at $\varphi=\pi$ and at a temperature of
	$18\,$mK (a) and $350\,$mK (b). (c) The extracted even charge parity state occupation
	$p_\textrm{even}$ as a function of the temperature. The inset shows the
	modulation of the spectrometer current at $2eV_\textrm{spec}=180\,\mu$eV at
	these two temperatures, which defines $\delta I_\textrm{odd}$ and $\delta
	I_\textrm{even}$, see text. The fit lines in (c) are based on Eq.~(3), without
	(blue dashed line) and including overheating (solid red line).}
\end{figure}

In conclusion, we performed broadband microwave spectroscopy on the gate charge
and phase-dependent energy dispersion of InAs/Al hybrid CPTs, utilizing an
on-chip nanofabricated circuit with a superconducting tunnel junction as a
frequency-tunable microwave source. We understand the observed spectra based on
the Hamiltonian of the circuit and find the characteristic charging and
Josephson tunneling energy scales, both exhibiting strong modulation with the
electrostatic gates coupled to the semiconductor channels. This broad tunability
demonstrates the feasibility of prospective topological qubits relying on a
controlled suppression of the charge modulation. Finally, we directly measure
the time-averaged even and odd charge parity occupation of the CPT island,
yielding a residual odd occupation probability and unpaired quasiparticle
density which are much higher than typical values acquired earlier for
all-metallic devices. This can be a limiting factor for topological quantum bit
architectures that rely on charge parity manipulation and readout.

The analyzed raw data sets and data processing scripts for this publication are
available at the 4TU.ResearchData repository \cite{rawdata}.

The authors gratefully acknowledge O.~Benningshof and R.~Schouten for technical
assistance as well as D.~J.~van Woerkom, D.~Bouman and B.~Nijholt for
fruitful discussions. This work was supported by the Netherlands Organization
for Scientific Research (NWO) as part of the Frontiers of Nanoscience program,
Microsoft Corporation Station Q, the Danish National Research Foundation and a
Synergy Grant of the European Research Council. P.~K.~acknowledge funding from
the European Research Council (ERC) under the grant agreement No.~716655
(HEMs-DAM).

\bibliography{spectroscopy_bibliography}

\end{document}